\documentclass[]{spie}

\usepackage{amsmath,amsfonts,amssymb}
\usepackage{graphicx}
\usepackage[colorlinks=true, allcolors=blue]{hyperref}
\usepackage{rotating}
\usepackage{threeparttable}
\usepackage{mathtools}

\title{Improving ALMA’s data processing efficiency using a holistic approach}

\author[a,e]{Theodoros Nakos}
\author[a]{Harold Francke}
\author[b,c]{Kouichiro Nakanishi}
\author[d]{Dirk Petry}
\author[d]{Thomas Stanke}
\author[e]{Catarina Ubach}
\author[a]{Luciano Cerrigone}
\author[e]{Erica Keller}
\author[g]{Alfonso Trejo}
\author[b]{Junko Ueda}

\affil[a]{Joint ALMA Observatory, Alonso de Cordova 3107, Santiago, Chile}
\affil[b]{National Astronomical Observatory of Japan, 2-21-1 Osawa, Mitaka, Japan}
\affil[c]{The Graduate University of Advanced studies, Sokendai, Japan}
\affil[d]{European Southern Observatory, Karl-Schwarzschild-Strasse 2, Garching, Germany}
\affil[e]{National Radio Astronomy Observatory, 520 Edgemont Rd, Charlottesville, USA}
\affil[g]{Academia Sinica Institute of Astronomy and Astrophysics, 11F of ASMAB, AS/NTU
No.1, Section 4, Roosevelt Road, Taipei, 10617, Taiwan}

\authorinfo{Further author information: (Send correspondence to Th. Nakos)\\ Th. Nakos: E-mail: theodoros.nakos@alma.cl,  H. Francke: E-mail: harold.francke@alma.cl, K. Nakanishi: E-mail: nakanisi.k@nao.ac.jp,  D. Petry: E-mail: dpetry@eso.org, T. Stanke: E-mail: tstanke@eso.org, C. Ubach: E-mail: cubach@nrao.edu, L. Cerrigone: E-mail: luciano.cerrigone@alma.cl, E. Keller: E-mail: ekeller@nrao.edu, A. Trejo: E-mail: trejo@asiaa.sinica.edu.tw, J. Ueda: E-mail: junko.ueda@nao.ac.jp}

\begin{document} 
\maketitle

\begin{abstract}
ALMA (Atacama Large Millimeter/submillimeter Array) is the world’s largest ground-based facility for observations in the millimeter/submillimeter regime. 
One of ALMA's outstanding characteristics is the large effort dedicated to the quality assurance (QA) of the calibrated and imaged data products offered to the astronomical community. The Data Management Group (DMG), in charge of the data processing, review, and delivery of the ALMA data, consists of approximately 60 experts in data reduction, from the ALMA Regional Centers (ARCs)  and the Joint ALMA Observatory (JAO), distributed in fourteen countries. With a throughput of more than 3,000 datasets per year, meeting the goal of delivering the pipeline-able data products within 30 days after data acquisition is a huge challenge.  

This paper presents (a) the history of data processing at ALMA, (b) the challenges our team had and is still facing, (c) the methodology followed to mitigate the operational risks, (d) the ongoing optimization initiatives, (e) the current data processing status, (f) the strategy which is being followed so that, in a few Cycles from now, a team of approximately 10 data reducers (DRs) at JAO can process and review some 80\% of the datasets collected during an observing cycle, and, finally, (g) the important role of the ARCs for processing the remaining datasets.  
\end{abstract}


\section{ALMA in a Nutshell}
\label{sec:intro} 

ALMA is a large international observatory that makes astronomical observations at millimeter and submillimeter wavelengths. ALMA's specified frequency range is 35 to 950 GHz, corresponding to an approximate wavelength range of  8.5 mm to 0.3 mm. Because these frequencies are strongly absorbed by clouds and water vapor in the atmosphere, the instrument is located in the very dry Atacama region of northern Chile. The actual location of the observatory itself is the Chajnantor plateau, which is at an altitude of 5000 meters above sea level. 

The instrument is made up of an array of parabolic antennas that work together to form a single telescope. Despite the antennas being quite similar in appearance compared to those used for lower-frequency radio astronomy and for communications, they are specially designed to be extremely stiff and to have a surface accuracy better than 12 microns while operating in an extremely hostile environment. There are 12 dishes of 7-meter diameter to form a close-packed array, 50 dishes of 12-meter diameter to form an extended array, and four dishes of 12-meter diameter that are used primarily as single-dish telescopes. The antennas in the 12-meter array can be moved around on the plateau to provide different configurations, ranging in extent from about 150 meters up to 16 kilometers. 

ALMA initiates a new observing cycle every October. The current cycle is Cycle 7  and it was expected to last until the end of September 2020. However, due to the COVID-19 contingency, it was decided to extend Cycle 7 until the end of September 2021. Thus, Cycle 8 will officially start on October 1st 2021. 

A new cycle does not only mark the start of the  acquisition of data coming from new proposals, but also the use of (1) a new ``on-line" software (i.e. a new version of all software components involved in data acquisition), (2) new observing capabilities not previously offered to the astronomical community (such as longer baselines, higher frequency observations, etc) and (3) a new pipeline, capable of processing both the old and possibly (even if partially) the new observing modes offered to the community with the new observing cycle.   

Most of the ALMA data are processed through the ALMA pipeline, which is based on CASA (Common Astronomy Software Applications)\footnote{For more information on CASA, see \url{http://casa.nrao.edu}}. Specific observing modes, such as solar, VLBI, and polarization, produce data that have to be processed manually (see details in Section~\ref{subsec:pl-vs-man}), rather than through the ALMA pipeline. Independently of how a dataset has been processed, though, it will go through a Quality Assurance process, to ensure that the data products that will be delivered to the Principal Investigators (PIs) meet the expected quality standards, in terms of sensitivity and angular resolution (AR), are free of observational and instrumental issues, and their corresponding calibration follows the observatory guidelines. 

In the case that the generated products comply with the PI's sensitivity and AR, they will be ingested to the ALMA Archive, setting a proprietary time of 12 or 6 months, depending on the type of the proposal. After that, the data become public to the astronomical community, through the ALMA science archive. 

\section{ALMA QUALITY ASSURANCE}
\label{sec:qa} 

The goal of ALMA Quality Assurance (QA)\index{Quality Assurance} is to ensure that the data products delivered to the PIs meet the expected quality standards. 
That is, the delivered products have reached the desired control parameters outlined in the science goals (or are as close to them as possible), they are calibrated to the desired accuracy, and calibration and imaging artifacts are mitigated as much as possible.  

To be more efficient in detecting problems, ALMA QA has been divided into several stages that mimic the main steps of the data flow. The broad classification\index{Quality Assurance!classification} of this multi-layered QA approach is: 

\begin{description}
\item[QA0:] The first check and monitoring of the calibration and array performance during and just after an observation. 
\item[QA1:] This involves the measurement of performance parameters and telescope properties by the observatory, which vary at the scale of months (or longer).
\item[QA2:] Full calibration and generation of imaging products and verification of compliance on PI science requirements of beam shape and image sensitivity.
\item[QA3:] The reporting, investigation and troubleshooting (if required) of issues found with the data by the PI or ALMA staff, after data delivery. 
\end{description}
A detailed description of the ALMA QA process can be found in chapter 11 of the ALMA Technical Handbook~\cite{remijan2019}.

\section{The Data Management Group}
\label{sec:dmg-intro} 

\subsection{General Description}
\label{sec:dmg-general}

The Data Management Group (DMG) is in charge of the processing, review and delivery of all ALMA data. Although organically DMG belongs to the Department of Science Operations (DSO) at JAO, and consists of some 9 experts in data processing (DP), who report directly to the DMG Manager, in its wider sense DMG comprises of DP experts at JAO, the ALMA Regional Centers (ARCs) and the ARC-nodes. 

The structure of the ARCs and ARC-nodes is as follows:
\begin{itemize}
    \item EA ARC : consists of the East Asian ALMA Regional Center (EA ARC) at the NAOJ Headquaters, Mitaka, Japan, and nodes at ASIAA, Taipei (Taiwan), and at KASI, Daejeon (Republic of Korea).
    \item EU ARC : consists of the European ALMA Regional Center (EU ARC) at the ESO Headquarters, Garching, Germany, and its system of nodes at  Ondrejov (Czech Republic), Grenoble (France), Bonn/Cologne (Germany), Bologna (Italy), Leiden (Netherlands), Onsala (Sweden), Manchester (UK), and an expertise centre in Lisbon (Portugal).  
    \item NA ARC : consists of the North American ALMA Regional Center (Charlottesville, VA, USA) and the Canadian ARC-node in Victoria (BC). 
\end{itemize}
 
The DMG Manager coordinates all DP-related activities with the help of the four Data Reduction Manager (DRM) leads at the ARCs (one per ARC) and JAO. The regional distribution of the DMG person power and the corresponding Full Time Equivalents (FTEs) is shown in Table~\ref{tab:dmg-pp-ftes}.

\begin{table}[ht]
\caption{DMG structure, person power and corresponding Full Time Equivalents at the beginning of Cycle 7. } 
\label{tab:dmg-pp-ftes}
\begin{center}       
\begin{tabular}{|l|c|c|} 
\hline
\rule[-1ex]{0pt}{3.5ex}  Site & $\#$ of Persons & $\#$ FTEs \\
\hline
\rule[-1ex]{0pt}{3.5ex}  JAO & 9  & 9  \\
\hline
\rule[-1ex]{0pt}{3.5ex}  EA ARC & 21 & 5  \\
\hline
\rule[-1ex]{0pt}{3.5ex}  EU ARC & 4 + 21 (at the nodes) & 2 + 3 \\
\hline
\rule[-1ex]{0pt}{3.5ex}  NA ARC & 10 & 5 \\
\hline 
\end{tabular}
\end{center}
\end{table}

\subsection{DMG's evolution}
\label{sec:dmg-history}

ALMA started Cycle~0 science observations on 30 September 2011. Over the course of nine years, ALMA has gone through eight observing Cycles, with over 13,900 datasets delivered (see Table~\ref{tab:Cycle_summary}).  Although going from delivering 320 datasets in Cycle~0 to delivering some 3013 datasets in Cycle~5 is impressive, getting to the level of the processing throughput required to keep up with the telescope’s observing efficiency took many years of learning, developing better ways of bookkeeping, consolidation of the team's experience in data processing and several years of optimization initiatives, across the ALMA partnership.

\subsubsection{Data processing} 
The QA2 process involves a complete, science-grade calibration of the ALMA data followed by high-quality imaging. If done purely manually, it requires highly-skilled data reducers (DRs) who will typically work intensively for several hours on an individual dataset (not counting breaks to wait for computing routines to complete). This cost in human resources was necessary at first, in order to better understand the quality of the data produced by the observatory, but it later turned into a tedious routine work, not sustainable on a long term. It was therefore planned from the start to develop a completely automated data processing pipeline.

This automated pipeline (short ``the Pipeline") was developed by first creating a sophisticated prototype, called script generator, which was used to create drafts of data processing scripts. Through this prototype, the DRs were able to sufficiently accelerate the routine part of the work. This turned out to be a good decision\cite{petry14}: in Cycle 3 the first half of the Pipeline (calibration only) was deployed in production. This pipeline calibrated the majority of the ALMA datasets, but the imaging had to be performed manually. In Cycle 4, a fully automated pipeline became available (calibration and imaging) — see Table~\ref{tab:casa_version} for CASA versions. It should be noted that the script generator is still of great use for developing reduction strategies for new modes (and to investigate corresponding heuristics), which are then transferred to the pipeline. 

Currently, more than 90\% of the ALMA datasets are processed with the ALMA pipeline.

\begin{sidewaystable}[ht]
\centering
\begin{threeparttable}[b]
\caption{A summary of the observation Cycles information. Observing range: provides the dates science observations were being conducted for a specific Cycle. Project ID: for projects submitted for an observing Cycle, the project code starts with the year the project was submitted for review. Beam size requirement: the amount the beam size value can be and still meet quality assurance 2 requirements, where ``A" is beam area and ``L" is beam length. For Cycle 7, where major and minor axes are ``a" and ``b” respectively, the beam area be between the areas determined by the minimum and maximum angular resolutions (assuming a 20~percent uncertainty. Sensitivity Requirement: the amount the sensitivity of the image can be and still meet quality assurance 2 requirements. For Cycles 0-2, the value presented applied to all observing bands. Delivered: the number of datasets delivered for each Cycle. Note: datasets could have been processed and delivered after the observing Cycle closed.} 
\label{tab:Cycle_summary}
\begin{tabular}{|l|l|l|l|l|l|} 
\hline
Cycle	&	Observing range	&	Project ID	&	Beam Size Requirement 	&	Sensitivity Requirements	&	Delivered 	\\
\hline											
0	&	30Sept2011-1Jan2013	&	2011	&	Best Efforts	&	40\%	&	320	\\
\hline											
1	&	1Jan2013-31May2014	&	2012	&	Best Efforts	&	15\%	&	631	\\
\hline											
2	&	3June2014-30Sep2015	&	2013	&	Best Efforts	&	10\%	&	1208	\\
\hline											
3	&	1Oct2015-30Sept2016	&	2015	&	A20\%	&	\begin{tabular}{@{}l@{}}10\% B3/4/6 \\ 15\% B7/8\\ 20\% B9 \end{tabular}	&	1938	\\
\hline											
4	&	1Oct2016-30Sept2017	&	2016	&	A69\%	&	\begin{tabular}{@{}l@{}} 10\% B3/4/6 \\ 15\% B7/8\\ 20\% B9/10 \end{tabular}	&	2325	\\
\hline											
5	&	Oct2017-30Sept2018	&	2017	&	L20\%	&	\begin{tabular}{@{}l@{}} 10\% B3/4/6 \\ 15\% B7/8\\ 20\% B9/10 \end{tabular}	&	3013	\\
\hline											
6	&	1Oct2018-30Sept2019	&	2018	&	L20\%	&	\begin{tabular}{@{}l@{}} 10\% B3/4/5/6 \\ 15\% B7/8 \\ 20\% B9/10 \end{tabular}	&	2849	\\
\hline											
7\tnote{A}	&	1Oct2019-30Sept2020	&	2019	&	minAR$<=sqrt(a*b)<=$maxAR	&	\begin{tabular}{@{}l@{}} 10\% B3/4/5/6 \\ 15\% B7/8\\ 20\% B9/10 \end{tabular}	&	1640	\\
\hline 
\end{tabular}
   \begin{tablenotes}
     \item[A]: Delivery totals up to $22^{nd}$ of March 2020. 
   \end{tablenotes}
 \end{threeparttable}
\end{sidewaystable}

\begin{sidewaystable}[ht]
\centering
 \begin{threeparttable}[b]
 \caption{A summary of the \textsc{casa} versions used for ALMA data processing in production, with the start and end use date in YYYY-MM-DD. \textit{Includes Pipeline} - if the \textsc{casa} version included the ALMA pipeline. The last 3 columns state which data processing workflow each \textsc{casa} version had available: Pipeline Calibration and Imaging (PipeCal+Img); Pipeline Calibration and Manual Imaging (PipeCal+ManImg); Manual Calibration and Imaging (ManCal+Img). nopipe = no pipeline included in \textsc{CASA} version; pipe = pipeline included in the \textsc{casa} version.}
 \label{tab:casa_version}
 \begin{tabular}{|l|l|l|l|l|l|l|} 
 \hline
CASA version	&	Use start	&	End Use	&	Includes Pipeline	&	PipeCal+Img	&	PipeCal+ManImg	&	Manual Cal+Img	\\
\hline
3.3.0	&	2012-04-01	&	2012-09-10	&	-	&	-	&	-	&	yes	\\
3.4.0	&	2012-09-10	&		&	-	&	-	&	-	&	yes	\\
4.2.0	&	2014-02-30	&	2014-05-14	&	-	&	-	&	-	&	yes	\\
4.2.1	&	2014-05-14	&	2014-09-22	&	-	&	-	&	-	&	yes	\\
4.2.2	&	2014-09-22	&	2015-08-07	&	yes	&	-	&	yes	&	yes	\\
4.3.0	&	2015-02-23	&	2015-03-09	&	-	&	-	&	-	&	yes	\\
4.3.1 nopipe	&	2015-03-09	&	2015-08-07	&	-	&	-	&	-	&	yes	\\
4.3.1 pipe	&	2015-08-07	&	2016-02-10	&	yes	&	-	&	yes	&	-	\\
4.4.0	&	2015-08-07	&	2016-04-08	&	-	&	-	&	-	&	yes	\\
4.5.1	&	2016-02-10	&	2016-04-01	&	yes	&	-	&	yes	&	-	\\
4.5.2	&	2016-04-01	&	2016-04-08	&	yes	&	-	&	yes	&	-	\\
4.5.3	&	2016-04-08	&	2016-10-06	&	yes	&	-	&	yes	& until 2016-04-20	\\
4.6.0	&	2016-04-20	&	2016-10-06	&	-	&	-	&	-	&	yes	\\
4.7.0	&	2016-10-06	&	2016-10-23	&	yes	&	yes	&	yes	&	yes	\\
4.7.0-1	&	2016-10-23	&	2017-03-31	&	yes	&	yes	&	yes	&	yes	\\
4.7.2	&	2017-03-31	&	2017-10-23	&	yes	&	yes	&	yes	&	yes	\\
5.1.1\tnote{A}	&	2017-10-23	&	2018-10-11	&	yes	&	yes	&	yes	&	yes	\\
5.3.0\tnote{B}	&	2018-05-01	&	2018-10-11	&	-	&	-	&	-	&	yes	\\
5.4.0-68\tnote{C}	&	2018-10-09	&	2019-02-10	&	yes	&	yes	&	yes	&	yes	\\
5.4.0-70	&	2019-02-11	&	2019-09-30	&	yes	&	yes	&	yes	&	yes	\\
5.6.1-8	&	2019-09-11	&	Present	&	yes	&	yes	&	yes	&	yes	\\
 \hline 
 \end{tabular}
   \begin{tablenotes}
     \item[A]: Not used for Solar datasets.
     \item[B]: Only used for Solar datasets. 
     \item[C]: A problem was found that affected mosaic images with limited number of integrations per pointing, in that the signal-to-noise in the first pointing (typically a corner) could be noticeably higher than in the rest of the image. Full details see the \textit{Release Notes for \textsc{CASA} 5.4.0} https://casa.nrao.edu/casadocs/casa-5.4.0/introduction/release-notes-540/
   \end{tablenotes}
 \end{threeparttable}
\end{sidewaystable}

\subsubsection{Analysis Tracking}
\label{subsec:ept}
During Cycle~0 all bookkeeping was maintained via \textsc{JIRA} tickets and the interface to the ALMA database (the Project Tracker). As the data output of the observatory picked up, this
process had to be further automated since increasing the number of personnel was not an option. Furthermore, the method used was not optimal for tracking more than a few hundreds of datasets. 
The automation was applied in two steps: for the next cycles, the lead Data Reduction Manager at the time, Eric Villard, developed a Python tool called ``Eric's Project Tracker" (EPT) which
automatically harvested information from specially formatted comments in \textsc{JIRA} tickets and thus permitted to monitor the ongoing analysis effort centrally. This required some discipline from the analysts but also gave full flexibility to introduce new features without waiting for long software release cycles.

The experience gathered with the EPT went into the design of the implementation of a set of QA tracking tools which were given the name ``ALMA Quality Assurance" (AQUA)~\cite{chavan16}. 
The first and most important step of the work towards AQUA  was to design the system of states where the different components of the ALMA data (Projects, Scheduling Blocks
\footnote{A Scheduling Block (SB) is the minimum set of instructions describing an ALMA observation that can be fully calibrated, such as (but not limited to), the positions and velocities of the science targets, details of the correlator setup and the integration and cycle times of the different calibrations. Detailed information about the SBs can be found in the ALMA Technical Handbook, at https://almascience.nrao.edu/documents-and-tools/cycle7/alma-technical-handbook/view.}, 
Execution Blocks
\footnote{An Execution Block (EB) is the result of a successful execution at the telescope of a SB. More information about the EBs can be found in the ALMA Technical Handbook.}, 
Member Observing Unit Sets
\footnote{A Member Observing Unit Set (MOUS) consists of one or more EBs. More information about the MOUSs can be found in the ALMA Technical Handbook.}
, etc.) can be in. Considering the different QA possibilities, depending on the component type, and the many exceptions and alternatives that needed to be taken into account, this ended up being a rather complex effort. In 2016, about half way through Cycle 3, and after being used only for QA0 purposes at first, AQUA finally also started to be used by the DRs. For some time it was used in parallel with its prototype, the EPT. Once AQUA was deemed mature enough, it replaced the EPT completely at the beginning of Cycle 5, in October 2017.

\subsubsection{QA2 judgement} 
During Cycle~0 a SB was observed until the estimated number of executions necessary to reach the sensitivity requested by the PIs was achieved. Then, the datasets were assigned to an ALMA scientist for data reduction (calibration and imaging) using CASA and Python. The data processing procedures were determined using Science Verification data, which were used to generate CASA  guides\footnote{http://casaguides.nrao.edu/index.php?title=ALMAguides}. If the calibration was successful, no instrumental artifacts were found in the images, the measured noise level was within 10\% to 20\%  of the requested value (depending on the frequency band), and the size of the synthesized beam was within the requested range, the datasets were classified as passing the second phase of quality assurance (QA2). 
These criteria underwent only moderate evolution and are documented in the ALMA Technical Handbook for each Cycle.

\subsubsection{Archive ingestions and early delivery rates} 
In Cycle 0, ALMA followed the path most observatories take. QA2 processing produced calibrated data which was ingested into the archive as Measurement Sets (MSs) in CASA table format. For a given ALMA raw dataset (ASDM), the corresponding calibrated MS has approximately twice the data volume. Consequently, ingesting the calibrated data in addition to the raw data, roughly tripled the archive size. In order to be economical, it was decided not to continue this practice in later Cycles and instead let the user download the raw data and restore the calibrated data products. 

Since the beginning of Cycle 1 the archive contains for each dataset the raw data, necessary calibration tables and scripts to generate the calibrated measurement set, the calibration logs, QA reports and the imaging products generated to assess QA2. 
After packaging and ingestion into the ALMA archive, the QA2 products were  made available to the PIs for download as a monolithic tarball together with individual tarballs for the raw data of each Execution Block (EB). The PIs were informed of a delivery via email through the ALMA Helpdesk. 

As of Cycle 5, all QA2 products are ingested in the archive as individual files, rather than tarballs. Furthermore, since March 2020, additional resources have been allocated to the re-ingestion in ASA, as individual files, of all data products from Cycles earlier than Cycle 5.

As far as the delivery rates are concerned, at the very beginning of ALMA operations the goal was to process the datasets within three working weeks, for a total delivery time (i.e. from data acquisition to delivery) of six weeks. Unfortunately, several issues arose, preventing this goal from being achieved, with the median total time for delivery being 84 days. This meant the observing cycle ended before all Cycle~0 datasets were delivered. This is no longer the situation, as it will be explained in more detail in subsection~\ref{subsec:lvl1kpi}.

\subsection{DMG's responsibility distribution}
\label{sec:dmg-distrib}
In the current scheme, DMG's responsibilities are structured as follows:
\begin{itemize}
    \item JAO: 
    \begin{itemize}
       \item Consolidate and optimize the data processing infrastructure at JAO. 
       \item Define Key Performance Indicators and Level 1 and Level 2 Metrics, which allow to measure DMG's performance in a consistent way on a monthly basis.
       \item Process 90\% of the datasets which go through the ALMA pipeline. This percentage might vary though, during certain periods (for example during the yearly maintenance of the JAO processing nodes).  
       \item Perform quality assessment at Execution Block level (QA0).
       \item Perform a quality assessment at MOUS level (QA2) of X\%  of the datasets processed at JAO, where ``X" varies as a function of Cycle. 
       \item Identify and execute optimization initiatives that can improve DMG's efficiency. These initiatives are related to (1) detecting problems with the data upstream in the process (i.e. just after data acquisition and before they reach DMG) and (2) minimizing the time spent on reviewing the data, with the final goal being that an algorithm does the final QA2 assessment and that the DRs only review the complex cases. 
    \end{itemize}
    \item ARCs:
    \begin{itemize}
       \item Process the remaining 10\% of the pipeline-able datasets, equally distributed among the three ARCs. This corresponds to roughly 2 datasets per week, per ARC.  
       \item Perform a quality assessment (QA2) of the remaining (100 - X\%) pipeline-able datasets, not reviewed by the DRs at JAO. 
       \item Manually process the ALMA data which cannot go through the ALMA pipeline (ca. 10\% of all ALMA data). Most of this data correspond to very specific observing modes such as solar, VLBI, polarization, etc. 
       \item Support the processing of forthcoming modes that will not yet be included in the pipeline (e.g. band-to-band phase transfer). 
    \end{itemize}    
\end{itemize}

\subsection{The DMG ecosystem}
\label{sec:ecosytem}

\begin{figure}[t]
 \centering
 \includegraphics[width=66mm]{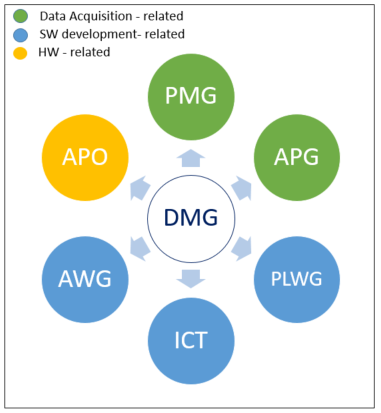}
 \caption{The DMG ecosystem: The Data Management Group interacts with the Program Management Group, the Array Performance Group, the Pipeline Working Group, the Integrated Computing Team, the Archive Working Group and the Archive and Pipeline Operations Group. The different colours indicate the main activity the group is focused on: green refers to activities related to data acquisition, blue for software development and orange when the main focus is on hardware.}
 \label{ecosystem}
\end{figure}

Because of the nature of DMG's work, our staff is involved in many more activities than just data processing and QA. The DMG ``ecosystem", i.e. the groups we are interacting with, are listed below:
\begin{itemize}
    \item Program Management Group (PMG), in charge of the data acquisition. The relation between PMG and DMG is bidirectional: PMG informs DMG about possible issues that might affect data processing and DMG gives feedback about the quality of the data processed, so as to avoid the execution of poor-quality observations.
    \item Array Performance Group (APG). APG is in charge of the data acquisition of data corresponding to new observing modes, using the data acquisition software (also known as ``on-line" software) that will be used for the forthcoming Cycle. DMG has to process these datasets, evaluate their quality, and report back to APG. 
    \item Pipeline Working Group (PLWG). DMG provides requirements to the PLWG regarding improvements or new features, identifies and reports issues, etc. Furthermore, approximately a month before the beginning of a new Observing Cycle, the PLWG delivers to DMG the version of the ALMA pipeline that will be installed in production at the beginning of the new Cycle and will be used to process the new-cycle ALMA data. DMG must run the pipeline both at JAO and the ARCs, and compare the results with a benchmark of approximately 10 -- 15 datasets (provided by the PLWG) to ensure that all sites obtain the same results. Once the validation has finished successfully, JAO officially accepts the new version of the pipeline, which is deployed in production and is used to process data of the new observing Cycle.   
    \item Integrated Computing Team (ICT). Updates of the ALMA software components not directly related to data acquisition go to production every two months (with some exceptions for specific months). Furthermore, there are on-going projects related to software development, which have a direct impact on data processing. DMG practically interacts with the ICT members (mainly, but not exclusively, with the developers and the release managers, both at JAO and the ARCs) on a daily basis, to create tickets for new features, coordinate the forthcoming activities (software deployment, advances of projects). Additionally, DMG performs integration testing of the features that will be included in the future software releases, in coordination with the ICT members. 
    \item All ALMA data which pass the QA2 criteria are ingested in the ALMA Science Archive (ASA). DMG is obviously the main internal (within ALMA) stakeholder to the Archive Working Group, which is in charge of the maintainability and evolution of the ASA.
    \item Archive and Pipeline Operations (APO) group, at JAO. APO is in charge of the processing infrastructure (processing nodes and disk space, the hardware infrastructure related to ASA and the associated data bases (among other tasks). DMG practically interacts on a daily basis with the APO colleagues, for discussing operational matters related to data processing, ingestion in the archive of the ALMA data, etc.
\end{itemize}
A schematic representation of DMG's relation with the above groups is presented in Fig.~\ref{ecosystem}. More details about some of the components of the DMG Ecosystem can be found in~\cite{espada14}.

\section{The Data Processing Workflow}
\label{sec:workflow}
This section is split in several subsections, each of them providing definitions which are relevant to the data processing workflow.

\subsection{The DMG deliverables: MOUS}
\label{mous}
The DMG deliverables to the PIs are the Member Observing Unit Sets. Every MOUS contains one or more Execution Blocks (details in subsection~\ref{subsec:ept}). 

The Level$-$2 Quality Assurance (QA2) is performed per MOUS. In other words, it is per MOUS that a decision will be taken about whether the data are a Pass (i.e. they comply with the angular resolution and sensitivity criteria) or Fail (they have failed at least one of them). 
 It should be stressed that one of the main motivations for a thorough QA2 analysis is to ensure that the PIs get what they ask for, meaning that an MOUS will go back to the observing queue if it does not meet the AR and sensitivity requirements (i.e. the MOUS is a QA2 Fail). In the case where no more data can be acquired, an MOUS will be delivered as SemiPass.

Every MOUS delivered to a PI (and which will become available to the astronomical community when the proprietary period is over) will contain (depending on whether it was processed by the pipeline or manually), continuum images, spectral line cubes, processing scripts, calibration tables and information related to the quality assurance (both at EB level and MOUS level) of the specific MOUS.   

\subsection{Pipeline-able versus manually processed MOUSs}
\label{subsec:pl-vs-man}
As previously mentioned, the MOUSs can be separated in two different categories regarding their processing type:
\begin{itemize}
    \item Pipeline-able: these MOUSs were successfully run with the ALMA pipeline and both calibration and imaging data products were generated. Depending on the quality of the images, it might be necessary to perform additional manual imaging, to fine-tune the input parameters of the CASA imaging tasks, and thus improve the quality of the image products. Nevertheless, as these MOUSs were initially processed with the ALMA pipeline, they are considered as pipeline-able.  
    \item Manually processed: these are MOUSs which can not yet be processed with the ALMA pipeline. They belong to specific observing modes or show some special features or problems that cannot be handled with the ALMA pipeline. The only way to reduce these MOUSs is by having a DR performing manual calibration and/or manual imaging. Both tasks are extremely time consuming and each can take between an FTE day to several FTE weeks of work prior to delivering the datasets, depending on the type of observations and/or the issue requiring manual intervention.  
\end{itemize}

\subsection{Weblog review}

For the MOUSs processed by the ALMA pipeline, one of the outputs is the so-called Weblog, which is a collection of information in HTML format (including several hundreds, or even more, graphs) related to the calibration and imaging of the specific dataset. 

The DRs both at JAO and the ARCs assess the quality of the data based on the information provided in the Weblog (although it might be necessary, for some cases, to look into more detail by extracting relevant information from the measurement sets).

\subsection{The Data Processing Workflow}
\label{subsec:workflow}
As expected, ALMA's data acquisition efficiency increased significantly as a function of observing cycle: from a few hundreds of datasets during Cycles 1$-$2, to a couple of thousand by the end of Cycle 4 (the exact numbers can be found in the last column of Table~\ref{tab:Cycle_summary}). 

To properly trace the status of the different MOUSs once they reached the hands of the DRs, DMG designed an automated system that helps to easily identify the part of the process (or ``state") at which every MOUS is found (also known as Data Processing Workflow, or DPW). Furthermore, based on the state of the MOUS, automated transitions could possibly take place, thus prioritizing in an optimized way the selection of the MOUSs to be dealt with by the DRs and DRMs. This workflow was placed in production at the beginning of Cycle 5 (October 2017), with several improvements taking place throughout the ALMA monthly software release cycles. 

Without going into many details, the DPW consists of the following MOUS states:
\begin{itemize}
    \item Fully Observed: an MOUS transitions to this state once the necessary number of Execution Blocks has been collected at the telescope. This transition takes place at the Operations Support Facility (OSF), as the MOUSs are initially saved in a temporary archive. %
    \item Ready for Processing: an MOUS transitions automatically to this state once the data have been transferred from the OSF to the ALMA Science Archive at JAO. 
    \item Processing: an MOUS shall transition to this state at the request of a data reducer. This transition is valid for all MOUS, regardless of whether they have to be processed manually or by the ALMA pipeline.  
    \item Ready For Review: this state transition ``declares" an MOUS as available to the DRs, for performing the QA2 assessment.  
    \item Reviewing: an MOUS shall transition to this state at the request of a data reducer. This transition is valid for all MOUS independently whether they have to be processed manually or by the ALMA pipeline. 
    \item Delivery in Progress: This is an automated transition, which takes place once the MOUS has been set to QA2 Pass or SemiPass.
    \item Delivered: This is an automated transition, which takes place once the MOUS has successfully been ingested in the Archive of the ``Preferred ARC" of the PI\footnote{The term ``Preferred ARC" refers to the ALMA Regional Center the PI is registered at (EA, EU or NA), during the proposal submission process. Every MOUS, regardless of where it was processed, must be ingested first in the ASA at JAO before it is replicated at the three archives at the ARCs.}.
\end{itemize}

The implementation of the DPW involves a suite of software components, which interact with the MOUSs and take specific actions, when necessary, according to the state in which the MOUS is found. The official terminology within ALMA for each software component involved in the operations of the Observatory is sub-system.

\section{DMG Key Performance Indicators (KPIs) and Metrics}
\label{sec:kpis}

\subsection{DMG's high-level objective and KPI}
\label{subsec:lvl1kpi}

The Level$-$1 KPI set by the Observatory to DMG is the following: 90\% of the pipeline-able MOUSs should be delivered within 30 days since they become Fully Observed (FO). This time-goal is relaxed to 45 days for the MOUSs that can only be manually processed. 

As explained in section~\ref{subsec:workflow}, an MOUS transitions to the FO state once all EBs needed to reach the associated sensitivity and AR goals have been collected. From that moment, the clock starts ticking for DMG, which is expected to deliver the specific MOUS within the expected time span.  

\begin{figure}[t]
 \centering
 \includegraphics[width=114mm]{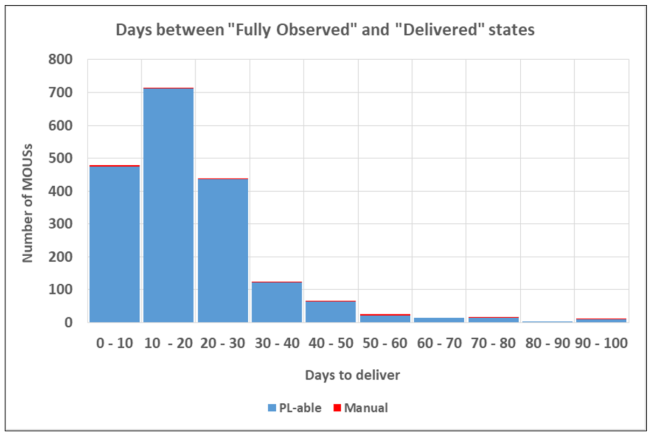}
 \caption{Example of the histogram used to monitor DMG's performance regarding the Level-1 KPI set by the observatory, according to which ``90\% of the MOUSs that go through the ALMA pipeline should be delivered within 30 days since they became Fully Observed." The specific histogram corresponds to MOUSs collected between October 1st 2019 and end of March 2020 (i.e. since the beginning of Cycle 7 and until the shutdown due to COVID-19) and which where delivered successfully. It contains information about some 1916 pipeline-able MOUSs (blue bars), out of which some 84\% (1616) were duly delivered, i.e. within 30 days. It also contains information about the 21 MOUSs that went through manual processing (red bars), out of which 13 (61\%) were delivered within 45 days.}
 \label{fig:histogram}
\end{figure}

The standard graph which allows to monitor DMG's performance regarding the Level-1 KPI is presented in Figure~\ref{fig:histogram}.

\subsection{Level 1, 2 and 3 metrics}
To reach its LVL$-$1 KPI, DMG, with the help of the ALMA Performance and Quality Manager (Bernhard Lopez), has defined a series of Level$-$1, 2 and 3 metrics, which allow to monitor its performance and take corrective actions, when necessary.  The classification of the metrics in three levels is based on the type and detail of information they contain, directly related to the profile of the recipients of the metrics (Observatory level, DMG level and DR level, for metrics LVL$-$1, 2 and 3, respectively). 

The time resolution of the LVL$-$1 and LVL$-$2 metrics has been chosen to be one month. The metrics are reset at the beginning of a new Cycle, i.e. every October, as ALMA operates with a new on-line (i.e. data acquisition) software and DMG processes the new Cycle data with a new pipeline. Cycle 7 will be the exception, as mentioned previously, due to its extension until end of September 2021.  

The time resolution of one month allows collecting enough statistical information (i.e. several hundreds of MOUSs) before drawing any conclusions. Other possibilities regarding the time resolution of the metrics (such as to associate them with the Array configurations) were rejected, as producing metrics on a monthly basis was aligned with the Observatory's reporting frequency.

Monitoring the metrics on a monthly basis, rather than at coarser time resolution, allows the team to react to anomalies in a pragmatic manner, and to check the results of recently-taken actions.
A typical example can be the deployment in production of a new software release, which brings features which are beneficial to DMG. By lowering the time resolution of the metrics (e.g. six months or more) it is possible that such useful information (i.e. the increment in efficiency due to the new features) is lost. Along the same line, a lower resolution might possibly translate into mixing bad performance with good performance data, thus not clearly identifying an inflexion point at the group's performance.   

Regarding the LVL$-$3 metrics, the time resolution will depend on the vision of the manager but also the DRM leads at the ARCs. As these metrics are related to the performance of the individuals, one must find the balance between avoiding micro-management (such as monitoring on a weekly basis how many data a DR has been working on) and not acting early enough to correct for issues which can affect DMG's performance.

\subsubsection{Detailed description of the DMG Level 1 metrics}
The Level$-$1 metrics allows the Observatory to monitor DMG's performance. More specifically, DMG reports to the ALMA management, on a monthly basis, on:
    \begin{itemize}
        \item The time evolution of the accumulated number of Fully Observed and Delivered MOUSs.
        \item The time evolution of the accumulated number of MOUSs which are ``Overdue" (i.e. MOUSs which have exceeded the 30/45 day time limit and have not been delivered yet).
        \item The time evolution of the number of MOUSs DMG is currently working on (also known as ``Workload").
        \item The time evolution of the accumulated number of MOUSs which have failed the QA2 criteria and the reasons for which they failed.
        \item The time evolution of the accumulated number of MOUSs which have gone to QA3 (see section~\ref{sec:qa}).   
    \end{itemize}

\subsubsection{General description of the DMG Level 2 and Level 3 metrics}
The Level$-$2 metrics are more detailed statistics used during the discussions between DMG and the teams forming the ``DMG Ecosystem" (see section~\ref{sec:ecosytem}), with the intention to trigger actions that will improve DMG's efficiency. 

\begin{figure}[t]
 \centering
 \includegraphics[width=134mm]{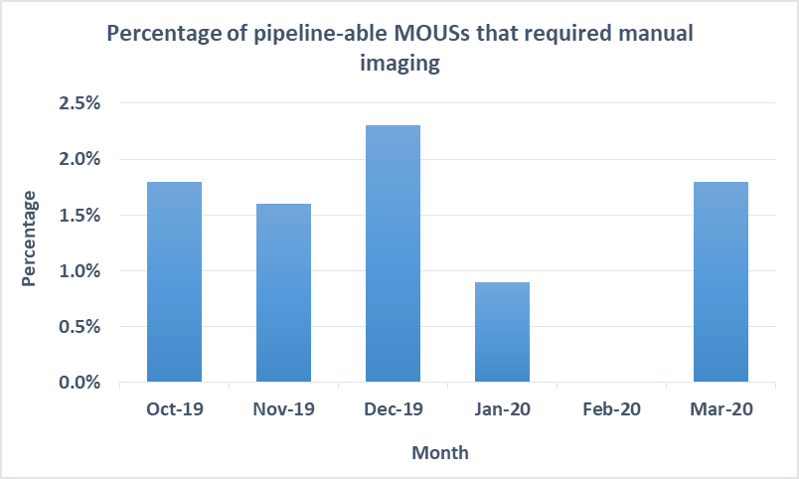}
 \caption{An example of a Level-2 metric. The specific metric presents the percentage of pipeline-able MOUSs, which transitioned to the Fully Observed state during a specific month, which needed additional manual imaging. February has no input as the Observatory shuts down for maintenance purposes, due to the non-optimal weather conditions.}
 \label{fig:pl-man-img}
\end{figure}

A representative example of a Level$-$2 metric is displayed in Fig.~\ref{fig:pl-man-img}. This specific graph shows the percentage of MOUSs, observed during a specific month, that needed additional manual imaging, after pipeline processing. The manual imaging is very expensive, as a data reducer must spend at least a couple of days' work to generate the final products (image cubes). This graph is shared with the relevant stakeholders (in this specific example with the Program Management Group and the Pipeline Working Group). By taking the necessary actions, this percentage is expected to drop in future Cycles.     

Level$-$3 metrics are inherent to the group, i.e. they are used to monitor the performance of the individual members at JAO and each of the ARCs and ARC-nodes. An example of a typical LVL$-$3 metrics is the number of MOUSs delivered per Observing Cycle, per data reducer. Given the granularity of the LVL$-$3 metrics, they can be re-visited at any time and be modified if needed. 

\section{Improving DMG's efficiency using a holistic approach}
\label{sec:actions}

After a careful mapping of the interactions of DMG with the different groups and sub-systems, at the beginning of Cycle 5 (October 2017), DMG followed a \textit{holistic} approach in order to improve its efficiency. This approach has been developed across seven axes, presented in the following sections. The effectiveness of the improvements will be checked via applicable KPIs.

\subsection{Consolidation of the hardware infrastructure at JAO}

The ALMA-wide (both JAO and ARCs) directive to DMG has been that JAO should be taking a higher processing and weblog review load as a function of Cycle, while the ARCs would mainly focus on the data that need manual processing. Several initiatives have allowed consolidating and maximising the performance of the cluster at JAO, which handles the processing of about 90\% of the pipeline-able data produced by ALMA as of Oct 2019 (aprox. 2700 MOUSs per year). Among these initiatives, we can mention:
\begin{itemize}
    \item Purchase of additional equipment (processing nodes, disk space).
    \item Redesign of the available disk space in smaller sectors with different characteristics/purposes, to match DMG's needs.
    \item Definitions of metrics to monitor the performance of the cluster.
    \item Optimization initiatives to maximize the cluster's throughput.
    \item Definition of policies about the cluster usage.
    \item Automation of the alerts.
    \item Regular coordination meetings, and 
    \item Knowledge transfer sessions.
\end{itemize}

\subsection{Improvement of communication and creation of awareness}    

 DMG has established regular meetings where all members of the ``DMG Ecosystem" are invited to. These meetings are the instance to:
 \begin{itemize}
     \item Report on DMG's performance.
     \item Communicate DMG's priorities.
     \item Create awareness, by showing how issues identified in the production environment influence DMG's efficiency. A typical case can be a bug in a software release not caught before the deployment. \textit{Creating awareness has been one of the key elements} in DMG's strategy since Cycle 5.  
     \item Bring together all sub-systems involved in data processing,  better coordinate their activities as a function of DMG's priorities and identify possible interconnections.
 \end{itemize}
 
Additionally, since Cycle 6 DMG has been reporting to the Observatory its LVL$-$1 KPIs on a monthly basis. A DMG daily report is also sent out observatory-wide. This report provides high-level information  about the most relevant data-processing metrics both for JAO and the ARCs.

\subsection{Minimization of the software threats}

Although the software cycle is a well-defined and controlled process within ALMA, there are instances that bugs were not caught during the verification and validation phase. Depending on the bug, the deployment in the production environment of the software feature containing a bug had an impact on DMG's performance which varied from (a) a few hours of not being able to process or review data, (b) to not being able to deliver data for a few days, or (c) up to having data reducers dedicating 100\% of their time for a full month on curating MOUSs whose metada were corrupted.   

Apart from creating awareness, through regular meetings and the respective analysis of the implications of the bugs with the colleagues involved in the software cycle, DMG has pursued the following additional approaches to minimize the software deployment risks:
\begin{itemize}
     \item Minimize the deployment in production of software patches in between software releases, through a better-defined filtering process.
     \item Improve the risk assessment process of the software patches that must imperatively go to production at any moment, instead of waiting for the next software deployment.
     \item Introduce the concept of integration testing. The end of any validation phase is followed by the so-called End-to-End tests (E2E), or integration tests. The E2E tests are tests which execute different use cases of the full data processing workflow (i.e. Fully Observed to Delivered) using a testing environment which simulates the current production environment plus the new features that will be deployed with the forthcoming software release.  
     \item At the beginning of a new observing cycle, and for approximately two weeks, give the highest priority to the processing and reviewing of the new-cycle datasets, rather than the older ones. Reverting the prioritization used throughout the rest of the cycle (oldest data have the highest priority) allows us to catch very fast possible issues with the new data acquisition software that goes to production at the beginning of a new Cycle.    
\end{itemize}

\subsection{Proper ticket reporting}  

ALMA uses the JIRA\MakeUppercase{\textregistered} platform for ticket reporting. DMG has given emphasis on the standardization of the procedures to follow as far as ticket reporting is concerned. This has been a challenging task, both because DMG interacts with so many sub-systems but also due to the geographical distribution of the staff and teams working for ALMA. However, clear instructions about the ticket reporting have payed off. A uniform way of reporting now gives visibility to the issues with higher impact on DMG's operations and consequently allows us to take actions based on a data-driven analysis.

\subsection{Creation of areas of expertise}  
Although by definition DMG is operating based on a distributed system, where experts at JAO and the ARCs perform all DP-related tasks, since Cycle 5 DMG has been promoting the creation of areas of expertise, by assigning specific tasks to given ARCs and ARC nodes, rather than following the scheme ``everybody does everything". 

Along this direction, JAO has been focusing on the processing and review of pipeline-able MOUSs exclusively. Several of its members are leading optimization initiatives which aim at maximizing DMG's efficiency in this area, by delivering the data faster and with less resources. 

Similarly, although all ARCs contribute to the processing and weblog review of the pipeline-able MOUSs, they have become the experts for the datasets which need manual calibration and manual imaging, tasks which are extremely time consuming and at the same time require a very high level of expertise.     

\subsection{Optimization initiatives}
\label{subsec:optim}
\begin{figure}[t]
 \centering
 \includegraphics[width=134mm]{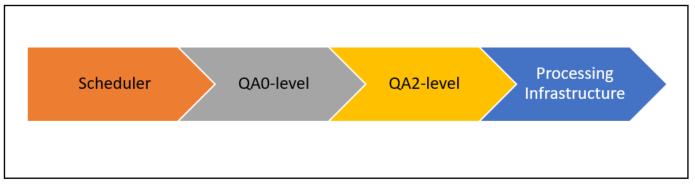}
 \caption{Level$-1$ description of the optimization initiatives that help increase DMG's performance, which include (1) the Scheduler sub-system, so that ALMA obtains the best possible data under the current array configuration, scheduling constraints and weather conditions (2) Several checks at QA0$-$level (i.e. per EB) so that problems are caught upstream (i.e. before reaching DMG), problematic data are flagged automatically and the observations are repeated, if necessary, (3) Optimization initiatives so that the DRs spent as little time as possible on QA2 assessment, and (4) Consolidation and optimization of the processing infrastructure at JAO, so that it can cope with the high demand to process the ALMA data and that the processing cluster's throughput is maximized.}
 \label{fig:optimizations}
\end{figure}

DMG members are engaged in several optimization initiatives, which are being carried out along four axes (also see Fig~\ref{fig:optimizations}:
\begin{itemize}
    \item Scheduling-related initiatives: Scheduler is the sub-system that schedules when and with what order the different Scheduling Blocks that will be executed at the telescope, to generate ALMA data (the output being EBs). The array configuration (i.e. distribution of ALMA antennas), the frequency at which the observations must be performed, the sensitivity and angular resolution requested by the PI and the current weather conditions (phase stability and precipitable water vapor, PWV) are some of the parameters that Scheduler uses as input when deciding which SB will be executed. These decisions have a direct impact on the quality of the data produced, thus a direct impact on the time that DMG must invest in the QA2 process of the produced MOUS. Furthermore, failure of an MOUS to comply with the QA2 requirements translates into requesting for more observing time for the same SB, thus impacting the completion of other projects. DMG members are actively involved in Scheduling-related initiatives with the aim to optimize the data acquisition$-$data processing chain.  
    \item QA0$-$level initiatives: The high-level requirement is to minimize the percentage of data that must be manually flagged\footnote{"Flag" is astronomy jargon for data cleaning, artifact removal, outlier removal, etc.} by a DR, when the MOUSs reach DMG, to less than 10\% of the Fully Observed MOUSs during an observing cycle. As the flagging happens at Execution Block (EB) level, the QA0$-$level software components, which operate per EB, must automatically flag these datasets. 
    \item QA2$-$level initiatives: Currently, the weblog of every pipeline-able MOUS is being reviewed by a DR, who performs the QA2 assessment. DMG has dedicated more than 1.5 FTEs during the last two years in initiatives which will automate the QA2 process by using a four-colour classification (also known as ``QA scores initiative"). The final goal is to have each task of the weblog consistently classified as green, blue, yellow or red, depending on the amount of effort that a DR must dedicate on the dataset prior to its delivery (with green meaning zero effort, thus it can be delivered without having a DR reviewing it, and red meaning very significant one).
    \item Processing infrastructure initiatives: Although the processing infrastructure at JAO is sufficiently big not to be considered a bottleneck regarding the delivery of the MOUSs with the established 30-day time span, there are cases where an a-priori estimation of the expected processing time of the MOUSs to enter the processing queue can be important. Producing such an estimation, with an acceptable uncertainty of the order of 10\%, has been an extremely challenging task until now, despite the different possibilities explored (type of Array, number of EBs, number of targets, baseline length, spectral resolution, etc).\footnote{Notably, an important variable in the processing time is the nature of the science target itself (extension in the sky and brightness) since that impacts the imaging time, and that is most the processing time of the whole process.} DMG, together with experts in industrial optimization processes from Adolfo Iba\~{n}ez university in Santiago (Chile) is engaged in a study whose aim is to generate a predictive model of the processing time of each MOUS. This information can be introduced as input to the software that optimizes the use of the resources of the processing cluster at JAO, thus optimizing its throughput. 
\end{itemize}
 
\subsection{Professional growth of the DMG members (at JAO)}    
One of the most common threats in a team's performance is the lack of motivation due to the repetition of the same tasks during long periods of time. In the case of DMG the task that falls into this category is the weblog review of the pipeline-able MOUSs, where a DR must check hundreds of plots to identify possible problems at QA2 level, i.e. after the processing has finished. After several Cycles and hundreds (or possibly more) of weblog reviews, for most of the DRs, this task does not present many challenges.  

As mentioned previously, the DRs at the ARCs and ARC-nodes are taking over the more challenging (therefore interesting) cases of the data which need manual processing, while the experts at JAO are mainly focused on reviewing the data that go through the pipeline. To avoid, on a long term, a possible ``burnout" and lack of motivation, they are all leading (or are actively being involved) in optimization initiatives (see the previous subsection), which do not only keep them intellectually stimulated but also gives them visibility as domain experts.   
Some of the DMG/JAO members (due to the limited spaces available, unfortunately) are being involved in activities related to management and/or leadership training, including a 360 evaluation, accompanied by a series of mentoring and/or coaching sessions. These activities are part of JAO's initiative in this area across all departments, which aims at generating a different culture within the organization, which finally translates into increased efficiency. The involvement of the DMG/JAO members in such activities does not only give them the tools for better interacting with colleagues, dealing with conflicts or managing projects but also a feeling of personal and professional growth.  

The involvement of the DMG/JAO members in the above initiatives has created a team fully engaged with the group and the organization, reaching unprecedented levels of excellence, similar to that of their DMG colleagues at the ARCs.

\section{DMG's strategic plan}
\label{sec:splan}
Upon a request from the ALMA management, the DMG manager presented in October 2019 the so-called DMG strategic plan. The plan consisted on (a) a thorough analysis that identified the most critical points that have an impact on DMG's performance, (b) a series of actions to be taken so as to optimize DMG's efficiency and (c) an estimate of the resources required so that 90\% of the pipeline-able MOUSs over an ALMA Cycle can be processed, reviewed and delivered from JAO. 

Without going into many details, the basic assumptions were (1) certain amount of hours on the sky for the different arrays (12$-$m, 7$-$m, TP), (2) both the remaining 10\% of the pipeline-able MOUSs and all MOUSs that need manual processing will be done at the ARCs and (3) that no-massive QA3 incident which will require reprocessing and review of hundreds of MOUSs takes place. 

\begin{figure}[t]
 \centering
 \includegraphics[width=134mm]{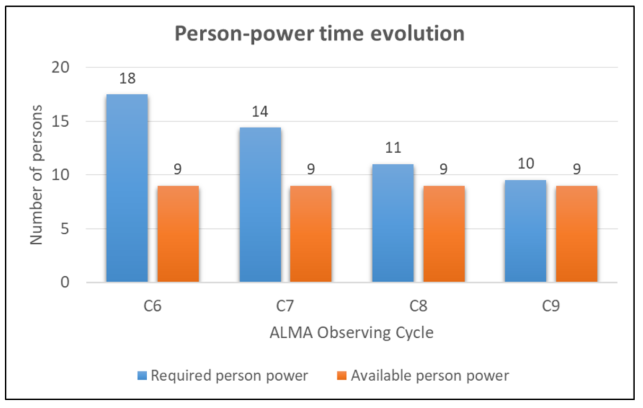}
 \caption{Time evolution of the number of DRs at JAO needed to review and deliver 90\% of the pipeline-able MOUSs produced during an ALMA Cycle. Currently, the difference between the requested person-power and the available person-power (9 DRs at JAO) is covered by the ARCs and ARC-nodes. The optimization initiatives which are currently taking place will gradually reduce the person-power to some 10 persons, by Cycle 9. Based on this projection, JAO will need to hire one DR to cope with the expected working load.}
 \label{fig:splan}
\end{figure}

Based on the optimization initiatives presented in subsection~\ref{subsec:optim}, and considering a certain percentage of improvements, per ALMA Cycle, it is expected that, by Cycle 9, with some 10 DRs (i.e. one more hire) JAO will be able to cope with the expected amount of work. The person-power time evolution of the DMG strategic plan is presented in Fig.~\ref{fig:splan}.            

\section{CONCLUSIONS}
After eight observing Cycles the Data Management Group has reached its peak performance, delivering about 85\% of the pipeline-able MOUSs within 30 days after data acquisition. A highly performing (fast, stable, intelligent) pipeline is our daily tool on which we trust the processing of some 2700 MOUSs on a yearly basis. A stable processing infrastructure is making sustainable the non-stop, 24/7, processing of the ALMA data throughout the year. The introduction of the Data Processing Workflow  allows DMG to precisely monitor the status of the incoming data. Several metrics, at three different levels, allow the group to both properly communicate to the relevant stakeholders about its own performance, but also to make corrections and improve the efficiency of the group. A series of optimization initiatives have brought major improvements on DMG's performance and, based on the estimations of DMG's strategic plan it will be possible to process and review 90\% of the pipeline-able datasets with 10 DRs at JAO by Cycle 9. The ARCs will be supporting JAO with the review of the remaining 10\% of the pipeline-able MOUSs and the processing of those which need manual calibration and/or imaging, a task that needs a very high level of expertise and a lot of dedication from the data reducers.          

\appendix    

\acknowledgments 
The authors of the paper would like to express their enormous gratitude to all past and current DMG members for their dedication and commitment to DMG's mission. Similarly, DMG would like to express its heartfelt thanks to all teams which form part of our Ecosystem and the individuals we are interacting with. Without their constant effort and professionalism we would have never reached this level of performance.   

ALMA is a partnership of ESO (representing its member states), NSF (USA) and NINS (Japan), together with NRC (Canada), MOST and ASIAA (Taiwan), and KASI (Republic of Korea), in cooperation with the Republic of Chile. The Joint ALMA Observatory is operated by ESO, AUI/NRAO and NAOJ.

\bibliography{report} 
\bibliographystyle{spiebib} 
\end{document}